\documentclass[prl, aps, twocolumn, amsmath, amssymb, 10pt]{revtex4-2}

\usepackage{amsmath,graphicx,color}
\usepackage{bm}
\usepackage{natbib}
\usepackage{placeins}
\setcitestyle{square}
\usepackage{lipsum}
\usepackage{braket}
\usepackage{leftindex}
\usepackage{makerobust}
\usepackage{xcolor}
\usepackage{hyperref}
\hypersetup{%
 colorlinks,
 breaklinks=true,
 plainpages=false,%
 citecolor=blue,
 linkcolor=blue,
 urlcolor=blue,
 bookmarksopen=true,%
 bookmarksnumbered=false,%
 bookmarksdepth=5%
}

\newcommand{\makeauthor}[2]{\newcommand{#1}[1]{{%
 \sffamily\color{#2}{%
 \bfseries\begingroup\escapechar=-1\edef\x{\endgroup\string#1}\x:%
 } ##1}}%
 \MakeRobustCommand#1}
\makeauthor{\dkm}{red}
\makeauthor{\al}{blue}
\makeauthor{\jb}{purple}

\begin{document}

\title{Josephson scanning tunneling spectroscopy in superconducting phases coexisting with pair-, charge- and spin-density-waves}

\author{Alyson Laskowski, Jasmin Bedow, and Dirk K. Morr \\
{\textit{Department of Physics, University of Illinois Chicago, Chicago, IL 60607, USA}} \\
\today
}

\begin{abstract}
We demonstrate that the recent observations of spatial oscillations in the energy position of the superconducting coherence peaks in the cuprate, transition metal dichalcogenide, iron-based and heavy-fermion superconductors are consistent with the possible presence of a pair-, charge- or spin-density-wave phase. We show that for all three cases, the spatial oscillations of the superconducting order parameter, $\Delta({\bf r})$ can be imaged via the critical Josephson current, $I_c({\bf r})$ measured in Josephson scanning tunneling spectroscopy experiments. Finally, we show that the spatial oscillations of the density waves and of $\Delta({\bf r})$ exhibit relative phase shifts of $\Delta \phi = 0$ or $\pi$ for a charge-density wave, and $\Delta \phi =\pi/2$ for the spin-density wave.
\end{abstract}

\maketitle


{\it Introduction} Identifying the nature of superconducting phases in correlated electron materials remains one of the most outstanding problems in unraveling their microscopic origin. This endeavor has been complicated by the fact that many of the materials of interest are known to have complex phase diagrams with intertwining charge/spin-density wave (C/SDW) and superconducting orders \cite{Fradkin2015}. Of particular interest in this regard are a series of scanning tunneling spectroscopy (STS) experiments in the cuprate superconductors \cite{Du2020, Hamidian2016}, the transition metal dichalcogenide 2H-NbSe$_2$ \cite{Cao2024, Liu2021}, the heavy-fermion UTe$_2$ \cite{Gu2023}, the iron-based  \cite{Kong2025,Liu2023,Zhao2023} and  kagome superconductors \cite{Deng2024,Chen2021,Han2025}, which have reported spatial oscillations in the energy of the superconducting coherence peaks, $E_c({\bf r})$, as observed in the measured differential conductance, $dI/dV$. These oscillations vary from a few to about 30 percent of the superconducting gap, and have been interpreted as a signature of an underlying pair-density wave (PDW) phase \cite{Agterberg2020a}. In this phase, the superconducting Cooper pairs possess a nonzero center-of-mass momentum, leading to spatial modulations in the superconducting order parameter $\Delta({\bf r})$ \cite{Agterberg2020a, Fradkin2015}, not unlike the Larkin-Ovchinikoff phase induced by large magnetic fields \cite{Larkin1965}. This interpretation received further support from measurements of the Josephson critical current, $I_c({\bf r})$ in Josephson scanning tunneling spectroscopy (JSTS) experiments \cite{Hamidian2016, Liu2021,Deng2024}. It was theoretically shown \cite{Graham2017a,Graham2019} that $I_c({\bf r})$ images the spatial form of $\Delta({\bf r})$, and can provide insight into its orbital symmetry \cite{Choubey2024}. Since the experimentally measured spatial oscillations in $I_c({\bf r})$ show the same wavelength as those of $E_c({\bf r})$ \cite{Liu2021,Deng2024}, it was argued that the latter are indeed a direct signature of a PDW.

The question naturally arises of whether the observed spatial oscillations in $E_c({\bf r})$ and $I_c({\bf r})$ indeed reflect a ``stand-alone" PDW, or whether they could also be induced by the coexistence of a charge- or spin-density-wave with a superconducting phase. This question was investigated in 2H-NbSe$_2$ \cite{Liu2021} and UTe$_2$ \cite{Gu2023}, where a CDW coexists with the superconducting state. While it was shown in both cases, that the wavelength of the oscillations in $E_c({\bf r})$ or $I_c({\bf r})$  is the same as that of the CDW, the spatial oscillations of the CDW and that of $E_c({\bf r})$ or $I_c({\bf r})$ possesses a relative phase of $\Delta \phi = 2\pi/3$ in 2H-NbSe$_2$ \cite{Liu2021} and of $\Delta \phi =\pi$  UTe$_2$ \cite{Gu2023}. The interpretation of all of these results is further complicated by the perplexing observation that in most of the studied materials, the wavelength of the spatial oscillations in $E_c({\bf r})$ is significantly smaller than \cite{Liu2021,Deng2024,Kong2025} or at most equal \cite{Du2020} to the superconducting coherence length, as it implies that superconducting properties can change on a lengthscale that is smaller than the size of a Cooper pair.

To provide theoretical insight into the origin of these experimental observations, in this article, we study the spectroscopic signatures of PDW, CDW or SDW phases coexisting with a superconducting phase with $s$-wave symmetry, which we refer to as PSC, CSC and SSC phases, respectively. In particular, we compare the features of these phases in the spatially and energy dependent local density of states (LDOS), $N({\bf r},\omega)$, as well as the spatially dependent Josephson current, $I_c({\bf r})$. We find that their signatures in $N({\bf r},\omega)$ exhibit a large degree of similarity, but vary significantly with the ratio between the oscillation wavelength, $\lambda$, (of the PDW, CDW or SDW) and the superconducting coherence length, $\xi$. While for $\lambda \gg \xi$, the spatial oscillations in the energy of the superconducting coherence peaks, $E_c({\bf r})$, mirror those of the superconducting order parameter, the oscillations in  $E_c({\bf r})$ decrease in amplitude and are more difficult to discern as $\lambda \rightarrow \xi$, resembling to a large degree the experimental observations in the differential conductance measured in STS experiments; these oscillations disappear as $\lambda$ becomes smaller than $\xi$.
Moreover, we demonstrate that for all three cases,  the spatially dependent Josephson current, $I_c({\bf r})$,  directly images the superconducting order parameter, $\Delta({\bf r})$, even when the oscillations in $E_c({\bf r})$ become difficult to identify. Finally, we show that the spatial oscillations of the CDW are either in phase with those of $I_c({\bf r})$, and hence $\Delta({\bf r})$, or are out-of-phase, the latter case being consistent with the experimental observation in UTe$_2$ \cite{Gu2023}. In contrast, the spatial oscillations in the SDW exhibit a phase difference of $\Delta \phi = \pi/2$ with those in $I_c({\bf r})$. Our results provide intriguing new insights into the possible mechanisms underlying the experimental STS ad JSTS observations.\\

{\it Theoretical Model} \
We begin by considering an $s$-wave superconductor coexisting with a PDW, CDW or SDW phase, as described by the Hamiltonian
\begin{align}
    \mathcal{H}_\mathrm{sc} &= -t\sum_{\langle \mathbf{r}, \mathbf{r}^\prime\rangle, \alpha} c^{\dagger}_{\mathbf{r},\alpha} c_{\mathbf{r}^\prime,\alpha}-\sum_{\mathbf{r},\alpha} \mu(\mathbf{r}) c^{\dagger}_{\mathbf{r},\alpha} c_{\mathbf{r},\alpha} \nonumber \\
    &-J\sum_{{\bf r}, \alpha, \beta} S({\bf r}) c^{\dagger}_{\mathbf{r},\alpha} \sigma^z_{\alpha, \beta} c_{\mathbf{r},\beta}-\sum_{\mathbf{r}}\left[\Delta(\mathbf{r}) c^{\dagger}_{\mathbf{r},\uparrow} c^{\dagger}_{\mathbf{r},\downarrow} + h.c.\right]
    \label{sc_ham}
\end{align}
where  $c^{\dagger}_{\mathbf{r},\alpha}$ $(c_{\mathbf{r},\alpha})$ creates (annihilates) an electron at site $\mathbf{r}$ with spin $\alpha$, $\sigma^z$ is the Pauli $z$-matrix, and $-t$ is the hopping amplitude between nearest neighbor sites on a two-dimensional square lattice. The emergence of a PDW on the background of an otherwise uniform $s$-wave superconducting order parameter (SCOP), which we refer to as a PSC phase, is described by a spatially varying  $\Delta(\mathbf{r})$ given by
\begin{align}
    \Delta(\mathbf{r}) = \Delta_0 + \Delta_\mathrm{PDW} \cos{\left(\mathbf{q}\cdot\mathbf{r}\right)} \ .
    \label{eq:sc_order_param}
\end{align}
Similarly, a CDW is described by a spatially varying chemical potential $\mu(\mathbf{r})$ of the form
\begin{align}
    \mu(\mathbf{r}) = \mu_0 + \Delta \mu \cos{\left(\mathbf{q}\cdot\mathbf{r}\right)}.
    \label{eq:chem_potential}
\end{align}
while an SDW is described by a spatially varying spin-polarization along the $z$-axis given by
\begin{align}
    S(\mathbf{r}) = S \cos{\left(\mathbf{q}\cdot\mathbf{r}\right)}.
    \label{eq:spin}
\end{align}
with $J$ being the exchange coupling between the spin-polarization, and the conduction electrons. For all cases considered, we assume that $\bf q$ in Eqs.~(\ref{eq:sc_order_param})-(\ref{eq:spin}) is directed along the $x$-axis. Below, we use $\mu_0=-3.6t$, yielding a superconducting coherence length $\xi=v_F/\Delta_0 \approx 25 a_0$, where $v_F$ is the Fermi velocity.

The presence of a CDW or SDW in the superconducting phase induces spatial oscillations in the SCOP, which are self-consistently computed using the BCS gap equation
\begin{align}
    \Delta(\mathbf{r}) = -\frac{V_0}{\pi}\int_{-\infty}^{\infty} \mathrm{d}\omega \; n_{F}(\omega) \text{Im}\left[F_s(\mathbf{r},\mathbf{r},\omega) \right] \ .
    \label{eq:self_consistent_eq}
\end{align}
Here, $V_0$ is the superconducting pairing potential, $n_{F}(\omega)$ is the Fermi distribution function, and $F_s(\mathbf{r},\mathbf{r},\omega)$ is the local retarded anomalous Green's function  of the superconductor. To calculate the latter, we rewrite Eq.(\ref{sc_ham}) as
\begin{align}
    \mathcal{H}_{\text{sc}} = \Psi^{\dagger} \hat{H}_\mathrm{sc} \Psi \ ,
    \label{}
\end{align}
where we introduced the spinor
\begin{align}
    \Psi^{\dagger} = \left(c_{1,\uparrow}^{\dagger}, c_{1,\downarrow}, ...,  c_{i,\uparrow}^{\dagger}, c_{i,\downarrow}, ..., c_{N,\uparrow}^{\dagger}, c_{N,\downarrow}\right)
    \label{eq:spinor}
\end{align}
with $i = 1, ..., N$ being an index to denote a site $\mathbf{r}$ in the system. The retarded Greens function matrix is then obtained via
\begin{align}
    \hat{G}_{\text{sc}}(\omega + \mathrm{i}\Gamma) = \left[\left(\omega + \mathrm{i}\Gamma \right)\hat{1} - \hat{H}_{\text{sc}} \right]^{-1}
    \label{eq:g_matrix}
\end{align}
where $\hat{1}$ is the $N \times N$ identity matrix and $\Gamma > 0$. The local anomalous Green's function at site $\mathbf{r}$ (index $i$), $F_{s}(\mathbf{r},\mathbf{r}, \omega)$, is the $(2i- 1, 2i)$ element of $\hat{G}_{\text{sc}}$. Similarly, we compute the local density of states, $N({\bf r}, \omega)$ using
\begin{align}
    N({\bf r}, \omega)= - \frac{1}{\pi} \sum_{\alpha} {\rm Im} \left[G({\bf r},{\bf r}, \alpha, \omega) \right]
\end{align}
where the spin resolved retarded Greens function $G({\bf r},{\bf r}, \alpha, \omega)$ is obtained from the diagonal elements of $\hat{G}_{\text{sc}}$.

Finally, the DC-Josephson current between an atomically sharp JSTS tip and a site $\mathbf{r}$ in the $s$-wave superconductor in the weak-tunneling limit, i.e., to lowest order in the tunneling amplitude $t_0$, is given by
\begin{align}
    I_{J}(\mathbf{r}) &= 8 \frac{e}{\hbar} t_{0}^{2} \sin{(\Delta\Phi)}\int\frac{d\omega}{2\pi}\; n_{F}(\omega) \text{Im}\left[F_s(\mathbf{r},\mathbf{r},\omega) F_{t}(\omega)\right] \nonumber \\
    &\equiv I_{c}(\mathbf{r}) \sin{(\Delta\Phi)} \ .
    \label{eq:josephson_current}
\end{align}
Here, $F_t$ is the retarded anomalous Green's function of the superconducting tip, $I_c$ is the critical Josephson current, and $\Delta\Phi$ is the phase difference between the superconducting order parameters of the tip and the $s$-wave superconductor. We take the tip anomalous Greens function to be that of a 2D bulk system, obtained similarly to $F_s(\mathbf{r},\mathbf{r},\omega)$, with a spatially constant superconducting order parameter $\Delta_{\text{tip}}$. We set $T = 0$ in all calculations. \\

{\it Results}
\begin{figure}[h]
 \centering
 \includegraphics[width=\columnwidth]{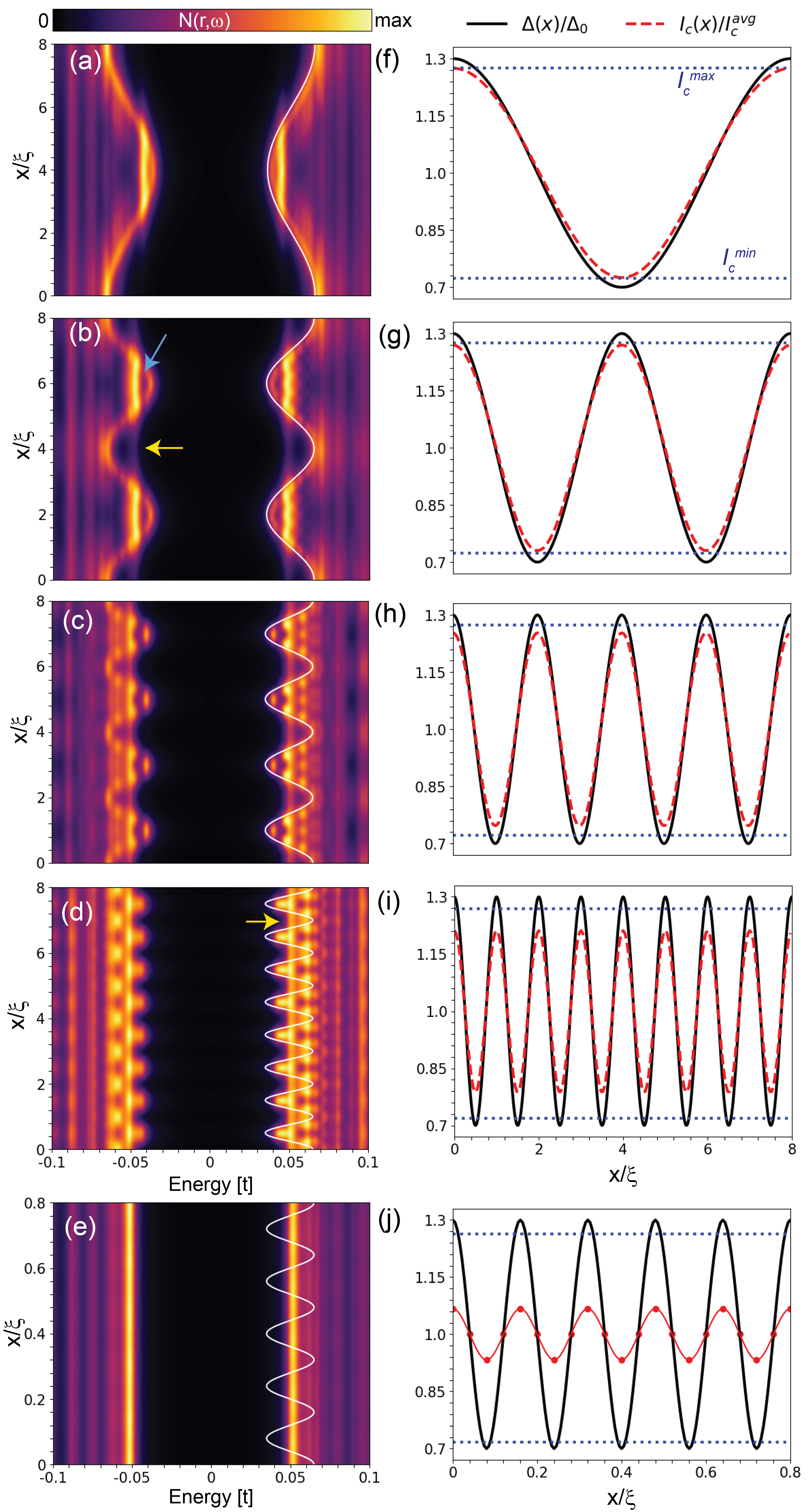}
 \caption{
    Spatial and energy dependence of the LDOS for a mixed PDW phase with four different wavelengths: (a) $\lambda = 200 a_0 = 8\xi$, (b) $\lambda = 100 a_0 = 4\xi$, (c) $\lambda = 50 a_0 = 2\xi$, (d) $\lambda = 25 a_0 = \xi$ and (e) $\lambda = 4 a_0 = 0.16\xi$. (f)-(j) show the spatial dependence of the normalized Josephson critical current $I_{c}(\mathbf{r})/I_{c}^{\text{avg}}$ with $\Delta_{\text{tip}} = \Delta_0$ (dashed red) and of $\Delta(\mathbf{r})/\Delta_{0}$ (solid black) for the corresponding cases in (a)-(e). Parameters are $(\Delta_0, \Delta_\mathrm{PDW}, \mu_0, \Delta \mu, JS) = (0.05, 0.015, -3.6, 0,0)t$. For (a)-(e), we set $\Gamma = 0.0025t$, while for (f)-(j) $\Gamma = 0.1t$.
 }
 \label{fig:Fig2}
\end{figure}
We begin by considering the spectroscopic signatures of a PSC phase, in which a spatial constant superconducting order parameter coexists with a spatially oscillating one, as described by Eq.(\ref{eq:sc_order_param}).
 To clearly identify the spectroscopic signatures of this phase, we take $\Delta_\mathrm{PDW}/\Delta_0=0.3$, consistent with the magnitude of the spatial oscillations observed in the iron-based superconductors \cite{Kong2025}. In Figs.~\ref{fig:Fig2} (a)-(e), we present the resulting spatial and energy-resolved LDOS, $N({\bf r},\omega)$ along ${\bf q}||{\hat x}$ for 5 different values of $\lambda/\xi$. For $\lambda = 200 a_0 =8 \xi$, the spatial variation in the energies of the coherence peaks, and thus of the superconducting gap, as reflected in $E_c({\bf r})$, smoothly follows that of the superconducting order parameter, $\Delta(\mathbf{r})$ (see white line). With decreasing $\lambda$ [see Fig.~\ref{fig:Fig2} (b) for $\lambda=100 a_0 = 4\xi$], we find that states that are localized at an energy $E_0$ near the minimum in $|\Delta(\mathbf{r})|$ (see blue arrow), begin to extend more and more into the ``forbidden" spatial regions where $|\Delta(\mathbf{r})|>E_0$ (see yellow arrows). This trend continues with decreasing $\lambda$ [see Fig.~\ref{fig:Fig2}(c) and (d) for $\lambda=50 a_0=2\xi$ and $\lambda=25 a_0 = \xi$, respectively], effectively filling in the LDOS in those regions where $|\Delta(\mathbf{r})|$ is the largest [see yellow arrow in Fig.~\ref{fig:Fig2}(d)]. As a result, the amplitude in the spatial variation of the superconducting gap, as observed in the LDOS, is smaller than that of the superconducting order parameter [cf. Figs.~\ref{fig:Fig2}(a) and (d)]. Finally, for $\lambda=4 a_0 \ll \xi$ [see Fig.~\ref{fig:Fig2}(e)], $N({\bf r},\omega)$ does not exhibit any discernible oscillations, consistent with the notion that superconducting properties can in general vary only on lengthscales that are larger than the size of Cooper pairs, i.e, the coherence length.

 Moreover, we find that the normalized Josephson critical current along in the direction of the oscillation direction images $\Delta(\mathbf{r})/\Delta_{0}$, although the agreement becomes worse with decreasing $\lambda$ [see Figs.~\ref{fig:Fig2}(f)-(j)]. The reason for this is that the local anomalous Greens function $F({\bf r},{\bf r},\omega)$ that contributes to the Josephson current [see Eq.(\ref{eq:josephson_current})] is determined by the form of the superconducting correlations in a region around ${\bf r}$ set by the coherence length $\xi$. As a result, when $\lambda$ approaches $\xi$ and the SCOP varies more rapidly, the agreement between $I_c({\bf r})$ and the (local) $\Delta({\bf r})$ diminishes. This interpretation is also supported when considering $I_c^{\text{max}}$ ($I_c^{\text{min}}$) as the value of $I_c({\bf r})$ for a homogeneous superconductor with a superconducting order parameter corresponding to the largest (smallest) value in the PSC phase: we see that as $\lambda$ decreases, the deviation of $I_c({\bf r})$ from $I_c^{\text{min},\text{max}}$ becomes increasingly larger.

\begin{figure}[h]
 \centering
 \includegraphics[width=\columnwidth]{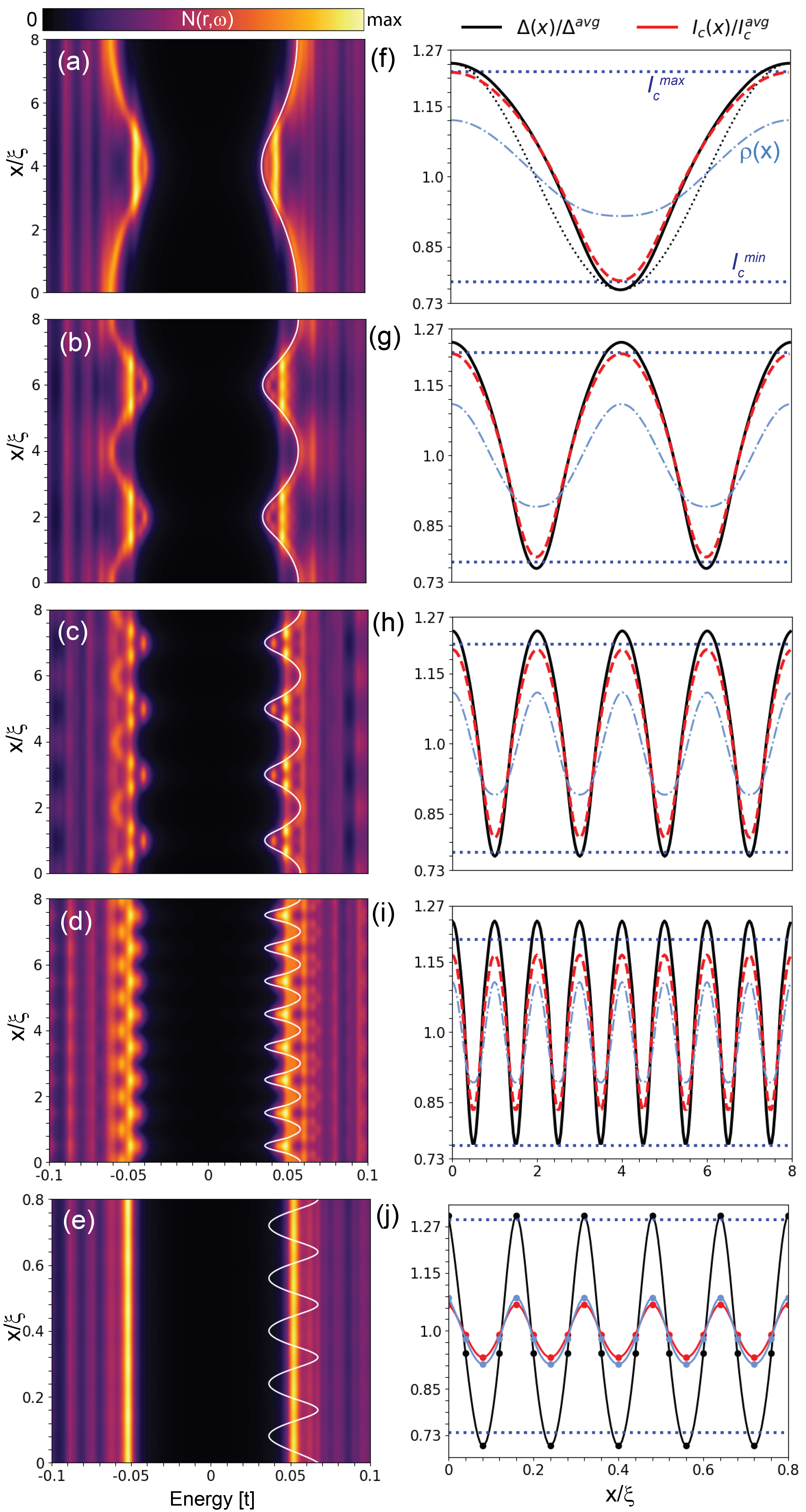}
 \caption{
    Spatial and energy dependence of the LDOS for a mixed CDW phase with four different wavelengths:(a) $\lambda = 200 a_0 = 8\xi$, (b) $\lambda = 100 a_0 = 4\xi$, (c) $\lambda = 50 a_0 = 2\xi$, (d) $\lambda = 25 a_0 = \xi$, and (e) $\lambda = 4 a_0 = 0.16\xi$. (f)-(j) show the spatial dependence of the normalized Josephson critical current $I_{c}(\mathbf{r})/I_{c}^{\text{avg}}$ with $\Delta_{\text{tip}} = \Delta_0$ (dashed red) and of $\Delta(\mathbf{r})/\Delta^{\text{avg}}$ (solid black) and charge density $\rho(\mathbf{r})$ (dashed blue) for the corresponding cases in (a)-(e). Parameters are $(\Delta_0, \Delta_\mathrm{PDW}, \mu_0, \Delta \mu, JS) = (0.05, 0, -3.6, 0.45, 0)t$. For (a)-(e), we set $\Gamma = 0.0025t$, while for (f)-(j) $\Gamma = 0.1t$.
 }
 \label{fig:Fig3}
\end{figure}
We next consider the case where a CDW coexists with the $s$-wave supercondctor, referred to as the CSC phase. The spatial variation in the chemical potential, $\mu({\bf r})$, which characterizes the CDW, induces spatial oscillations in the SCOP, $\Delta({\bf r})$, which are obtained from a self-consistent solution of the BCS gap equation Eq.(\ref{eq:self_consistent_eq}). This is reflected in a plot of the resulting LDOS $N({\bf r}, \omega)$ along ${\bf q}$ which exhibits spatial variations in the position of the coherence peaks, as shown in Fig.~\ref{fig:Fig3}(a)-(e) for different wavelengths $\lambda$ of the oscillations in $\mu$. Here, the value of $\Delta \mu$ was chosen such that the amplitude of the induced variation in $\Delta(\mathbf{r})$ is approximately the same as in the PSC phase, as shown in Fig.~\ref{fig:Fig2}. As a result, the LDOS in the CSC phase looks quite similar to that in the PSC phase, and we again find that once the CDW oscillation wavelength becomes smaller than $\xi$ [see Fig.~\ref{fig:Fig3}(e)], the spatial oscillations in the LDOS $N({\bf r},\omega)$ rapidly vanish.

Moreover, the spatial form of the normalized Josephson current very well images that of the superconducting order parameter, as shown in Figs.~\ref{fig:Fig3}(f)-(j). In addition, a plot of the charge density $\rho({\bf r})$ in Figs.~\ref{fig:Fig3}(f)-(j), reveals that the spatial oscillations in $\Delta(\mathbf{r})$ and $\rho({\bf r})$ possess the same wavelength, in agreement with the observations in Refs.\cite{Liu2021,Gu2023}. While $\Delta(\mathbf{r})$ and $\rho({\bf r})$ are in-phase in the results of Fig.~\ref{fig:Fig3}, the $\pi$-phase shift between $\Delta(\mathbf{r})$ and $\rho({\bf r})$ observed in Ref.\cite{Gu2023} can be reproduced when considering a more than half-filled band, as shown in Fig.~\ref{fig:Fig3a}, rather than the less than half-filled band considered in Fig.~\ref{fig:Fig3}.
\begin{figure}[h]
 \centering
 \includegraphics[width=\columnwidth]{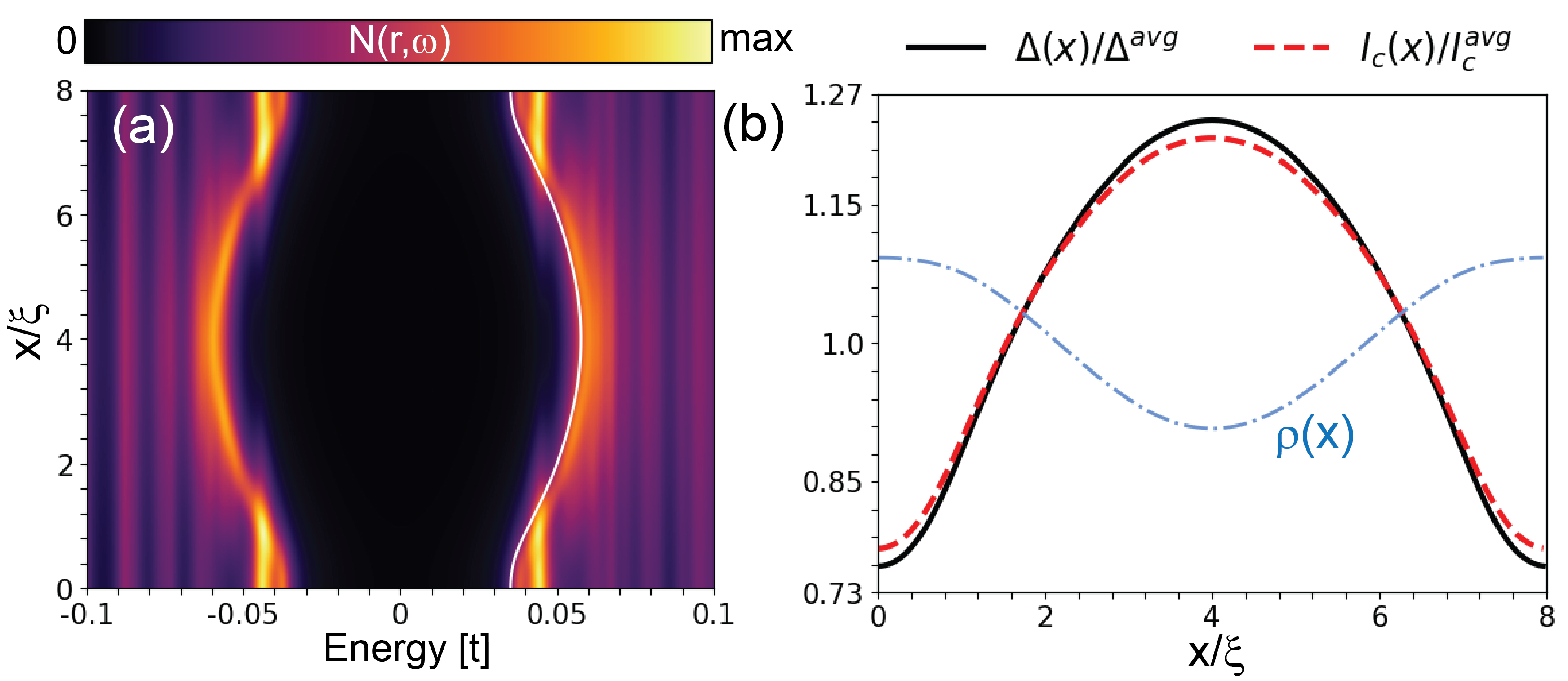}
 \caption{
    (a) Spatial and energy dependence of the LDOS for a CSC phase with $\lambda = 200 a_0 = 8\xi$, for a more than half-filled electronic band. (b) Spatial dependence of the normalized Josephson critical current $I_{c}(\mathbf{r})/I_{c}^{\text{avg}}$ with $\Delta_{\text{tip}} = \Delta_0$ (dashed red) and of $\Delta(\mathbf{r})/\Delta^{\text{avg}}$ (solid black) and $\rho(\mathbf{r})$ (dashed blue), corresponding to (a). Parameters are $(\Delta_0, \Delta_\mathrm{PDW}, \mu_0, \Delta \mu, JS) = (0.05, 0, 3.6, 0.45, 0)t$. For (a), we set $\Gamma = 0.0025t$, while for (b) $\Gamma = 0.1t$.
 }
 \label{fig:Fig3a}
\end{figure}
While the spatial dependence of the LDOS and that of the $I_c({\bf r})$ shown in Figs.~\ref{fig:Fig3} and \ref{fig:Fig3a} for the CSC phase look qualitatively very similar to those of PSC phase shown in Fig.~\ref{fig:Fig2}, there exists a small, but significant difference. In particular, while the spatial dependence of  $\rho({\bf r})$ is described by a cosine function [see blue lines in Figs.~\ref{fig:Fig3}(f)-(j)], the resulting spatial dependence of $\Delta({\bf r})$, and thus of $I_c({\bf r})$, deviates from that of a cosine function, as shown in Fig.~\ref{fig:Fig3}(f) by the black dotted line. We thus conclude that while the spatial oscillations in the LDOS and the SCOP induced by a CDW are similar to that in the PSC phase, they exhibit deviations from a pure cosine function, which can be detected via the Josephson current $I_c({\bf r})$.

Finally, we consider the effects of an SDW, as described by a spatially varying spin polarization $S({\bf r})$ [see Eq.(\ref{eq:spin})], on the spatial form of the superconducting order parameter, and thus of the resulting LDOS and Josephson current. Similarly to the CSC phase, the presence of an SDW induces spatial oscillations in the superconducting order parameter, $\Delta({\bf r})$, which we obtain from the self-consistent solution of  Eq.~(\ref{eq:self_consistent_eq}). The resulting spatial and energy-dependent LDOS for 5 different values of $\lambda$ is presented in Figs.~\ref{fig:Fig4}(a)-(e), which shows a similar evolution as for the PSC and CSC cases. In contrast to the case of a CSC, we now find that the wavelength of the induced oscillations in $\Delta({\bf r})$, $\lambda_{\Delta}$, is only half of that in $S({\bf r})$. The reason for this halving of the wavelength is that due to the spin-singlet nature of the superconductor, its response to a local spin polarization is invariant under $S({\bf r}) \rightarrow -S({\bf r})$. This is particularly apparent when comparing the spatial dependence of the Josephson current, $I_c({\bf r})$ [which again images well that of $\Delta({\bf r})$] with that of $S({\bf r})$ as shown in Figs.~\ref{fig:Fig4}(f) and (g). These plots reveal that any non-zero $S({\bf r})$ leads to a suppression of the SCOP, implying that the largest SCOP occurs at those spatial positions where $S({\bf r})=0$. This results in a phase shift of $\pi/2$ between the SDW, $S({\bf r})$, and $I_c({\bf r})$ [and hence $\Delta({\bf r})$]. Moreover, we note that the spatial form of the SCOP deviates from a cosine function [see dotted black line in Fig.~\ref{fig:Fig4}(f)], just as in the CSC case, and that the amplitude of the induced oscillations in $\Delta(\mathbf{r})$ strongly decreases with decreasing $\lambda_{\Delta}$. For example, in Fig. \ref{fig:Fig4}(g) ($\lambda_{\Delta} = 50 \; a_0 = 2\xi$), there is a $\sim 30\%$ variation in $\Delta(\mathbf{r})$, while in Figs. \ref{fig:Fig4}(h) and (i), which have $\lambda_{\Delta} = 25 \; a_0 =\xi$ and $12.5 \; a_0 = 0.5\xi$, respectively, the variations are only about $20\%$ and $12\%$. Once the wavelength of the SDW becomes smaller than the coherence length, we again find that any oscillations in the LDOS and SCOP are rapidly suppressed, as shown in Figs.~\ref{fig:Fig4}(e) and (j).

\begin{figure}[h]
 \centering
 \includegraphics[width=\columnwidth]{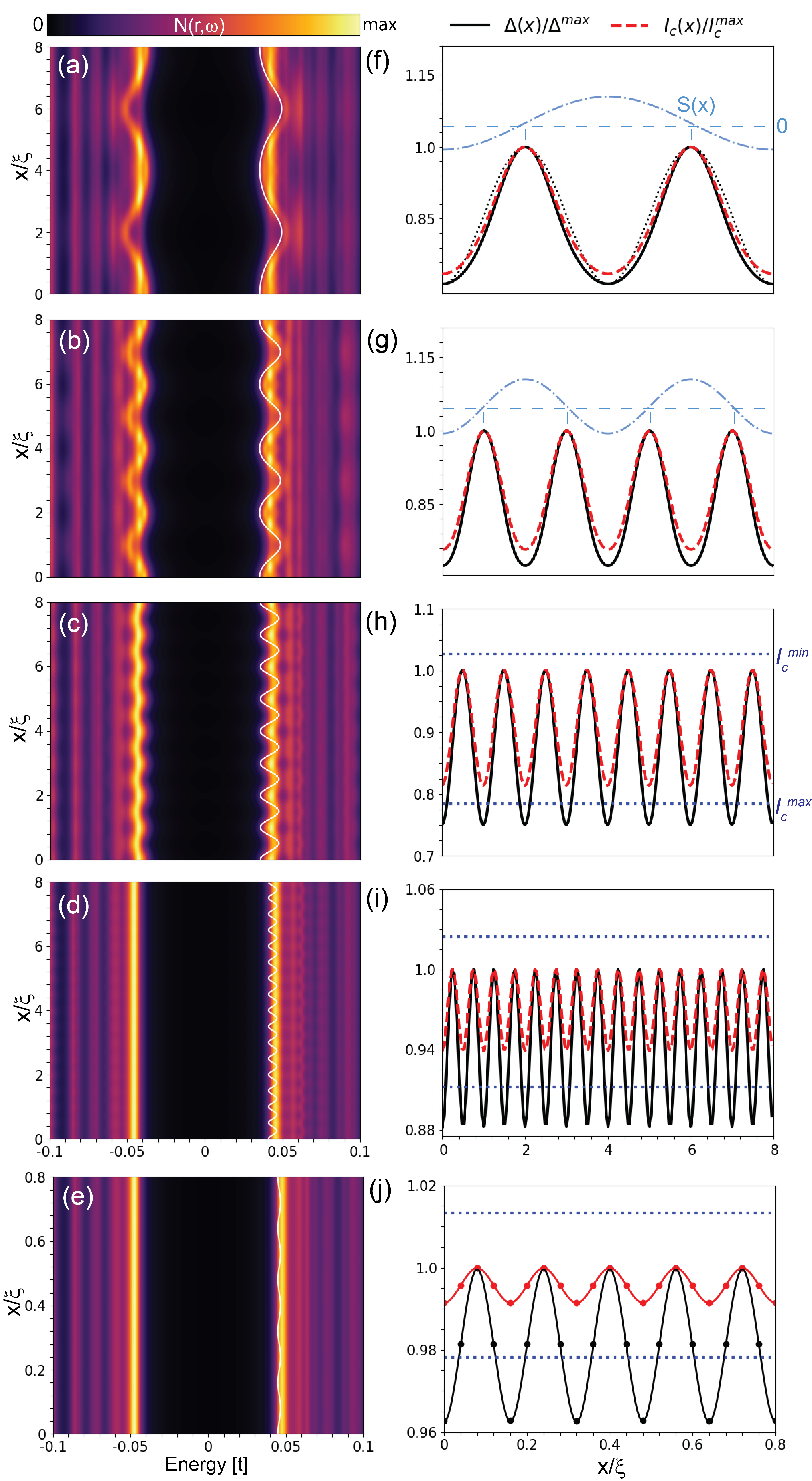}
 \caption{
    Spatial and energy dependence of the LDOS for a mixed SDW phase with four different wavelengths: (a) $\lambda = 200 a_0 = 8\xi$, (b) $\lambda = 100 a_0 = 4\xi$, (c) $\lambda = 50 a_0 = 2\xi$, (d) $\lambda = 25 a_0 = \xi$, and (e) $\lambda = 8 a_0 = 0.32\xi$. (f)-(j) show the spatial dependence of the normalized Josephson critical current $I_{c}(\mathbf{r})/I_{c}^{\text{max}}$ with $\Delta_{\text{tip}} = \Delta_0$ (dashed red) and of $\Delta(\mathbf{r})/\Delta^{\text{max}}$ (solid black) for the corresponding cases in (a)-(e). The spatial dependence of the spin polarization $S(\mathbf{r})$ (dashed blue) is additionally shown in (f) and (g). Parameters are $(\Delta_0, \Delta_\mathrm{PDW}, \mu_0, \Delta \mu, JS) = (0.05, 0.0, -3.6, 0.0, 0.2)t$. For (a)-(e), we set $\Gamma = 0.0025t$, while for (f)-(j) $\Gamma = 0.1t$.
 }
 \label{fig:Fig4}
\end{figure}

Finally, in Fig.~\ref{fig:Fig5}, we compare the spatially and energy-dependent LDOS in the PSC [Fig.~\ref{fig:Fig5}(a)], CSC [Fig.~\ref{fig:Fig5}(b)] and SSC phases [Fig.~\ref{fig:Fig5}(c)] with the experimental results for $dI/dV$ obtained in FeTe$_{0.55}$Se$_{0.45}$ \cite{Kong2025}[Fig.~\ref{fig:Fig5}(d)]. We find that in  all three phases considered above, the spatial oscillations in the LDOS qualitatively, and to a large extent quantitatively agree with the experimentally measured $dI/dV$.
\begin{figure}[h]
 \centering
 \includegraphics[width=\columnwidth]{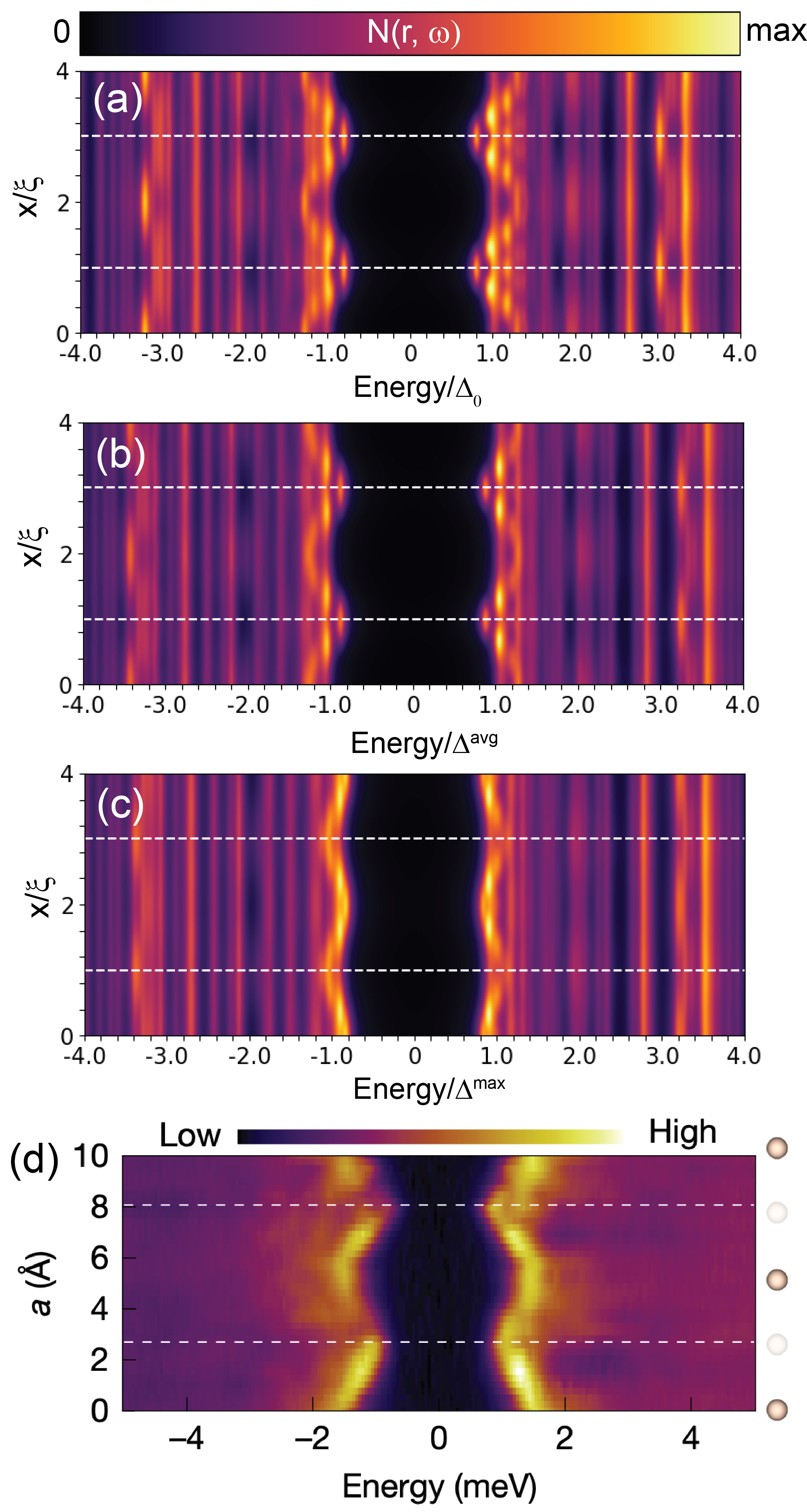}
 \caption{
    (a) Mixed PDW for wavelength $\lambda = 50 a_0 = 2\xi$ with same parameters as Fig.~\ref{fig:Fig2}.
    (b) Mixed CDW for wavelength $\lambda = 50 a_0 = 2\xi$ with same parameters as Fig.~\ref{fig:Fig3}.
    (c) Mixed SDW for wavelength $\lambda = 100 a_0 = 4\xi$ with same parameters as Fig.~\ref{fig:Fig4}.
    (d) Figure 2(g) of Ref.\cite{Kong2025}.
 }
 \label{fig:Fig5}
\end{figure}
We thus conclude that a measurement of the conventional $dI/dV$ is not sufficient to distinguish the microscopic origin of the observed oscillations in the LDOS. Rather, it is necessary to combine conventional $dI/dV$ measurements with those of the Josephson current, which exhibits some differences between the three phases, as well as with local spectroscopic techniques that directly probe  spatial oscillations in the charge- or spin-density, such as scanning quantum dot spectroscopy \cite{Wagner2019a} and spin-resolved scanning tunneling spectroscopy \cite{Wiesendanger2009a}, respectively. \\

{\it Conclusions} We demonstrated that the coexistence of a PDW, a CDW or an SDW with $s$-wave superconductivity, referred to as the PSC, CSC, and SSC phases, leads to spatial oscillations in the superconducting order parameter, $\Delta({\bf r})$. These, in turn, are reflected in spatial oscillations of (i) the energy position of the coherence peaks, $E_c({\bf r})$, as evidenced in the spatial and energy-resolved LDOS, $N({\bf r},\omega)$, and (ii) the critical Josephson current, $I_c({\bf r})$. In particular, we find that for all three phases, the spatial form of $I_c({\bf r})$, which can be measured via Josephson scanning tunneling spectroscopy, images that of the superconducting order parameter. Moreover, we have shown that the induced spatial oscillations in $N({\bf r},\omega)$ become weaker as the wavelength of the respective density waves, $\lambda$, approaches the superconducting coherence length $\xi$ from above, and are quickly suppressed when $\lambda$ becomes smaller than $\xi$. This reflects that superconducting properties do not vary on lengthscales smaller than the size of a Cooper pair.  Moreover, we show that the wavelength of the induced oscillations in $\Delta({\bf r})$ in the CSC phase is the same as that of the charge density, $\rho({\bf r})$, as observed in 2H-NbSe$_2$ \cite{Liu2021} and UTe$_2$ \cite{Gu2023}, but that the spatial oscillations of $\Delta({\bf r})$ and $\rho({\bf r})$ can be either in-phase ($\Delta \phi =0$) or out-of-phase ($\Delta \phi =\pi$), depending on the filling-factor of the electronic band. In contrast, in the SSC phase, the wavelength of the induced oscillations in $\Delta({\bf r})$ is only half of that of the spin polarization $S({\bf r})$, with the oscillations exhibiting a phase shift of $\Delta \phi =\pi/2$. Finally, a comparison of the theoretically computed LDOS $N({\bf r},\omega)$ with the experimentally measured differential conductance, $dI/dV$, shows remarkable agreement in all three PSC, CSC, and SSC phases. This implies that the experimental observations of spatial oscillations in $E_c({\bf r})$ and $I_c({\bf r})$ by themselves are insufficient to uniquely determine their microscopic origin, and thus need to be combined with other local spectroscopic techniques that directly probe  spatial oscillations in the charge- or spin-density, such as scanning quantum dot spectroscopy \cite{Wagner2019a} and spin-resolved scanning tunneling spectroscopy \cite{Wiesendanger2009a}, respectively.

Our results have raised a series of questions that need to be addressed in the future. Of particular interest is here the question of how the wavelength of the spatial oscillations observed in $E_c({\bf r})$ in several systems \cite{Liu2021,Deng2024,Kong2025} can be significantly smaller than their respective superconducting coherence lengths. One possible explanation could focus on the emergence of multi-band superconductivity, with the oscillating part of  $\Delta({\bf r})$ arising from bands with a much smaller coherence length than the superconducting bulk coherence length. Of similar interest is to understand the observed phase shift of $\Delta \phi =2\pi/3$ in 2H-NbSe$_2$ \cite{Liu2021} between the spatial oscillations in the charge density and the superconducting pair density, despite the oscillations exhibiting the same wavelength.  Future work is clearly required to address these open questions.\\

{\bf Acknowledgments}
This work was supported by the U.\ S.\ Department of Energy, Office of Science, Basic Energy Sciences, under Award No.\ DE-FG02-05ER46225. We would like to acknowledge helpful discussions with P. Hirschfeld.\\

{\bf Data availability}
Original data are available at (insert link to Zenodo depository).\\

{\bf Code availability}
The codes that were employed in this study are available from the authors on reasonable request.\\

\maketitle

\end{document}